# Topological Lifshitz transition and one-dimensional Weyl mode in HfTe$_5$

Wenbin Wu[1†], Zeping Shi[1†], Yuhan Du[1], Yuxiang Wang[2], Fang Qin[3], Xianghao Meng[1], Binglin Liu[1], Yuanji Ma[1], Zhongbo Yan[4], Mykhaylo Ozerov[5], Cheng Zhang[2,6*], Hai-Zhou Lu[3,7], Junhao Chu[8,9,10] Xiang Yuan[1,8*]

[1]State Key Laboratory of Precision Spectroscopy, East China Normal University, Shanghai 200241, China
[2]State Key Laboratory of Surface Physics and Institute for Nanoelectronic Devices and Quantum Computing, Fudan University, Shanghai 200433, China
[3]Shenzhen Institute for Quantum Science and Engineering and Department of Physics, Southern University of Science and Technology (SUSTech), Shenzhen 518055, China.
[4]School of Physics, Sun Yat-Sen University, Guangzhou 510275, China
[5]National High Magnetic Field Laboratory, Florida State University, Tallahassee, Florida 32310, USA
[6]Zhangjiang Fudan International Innovation Center, Fudan University, Shanghai 201210, China
[7]Shenzhen Key Laboratory of Quantum Science and Engineering, Shenzhen 518055, China
[8]School of Physics and Electronic Science, East China Normal University, Shanghai 200241, China
[9]Key Laboratory of Polar Materials and Devices, Ministry of Education, East China Normal University, Shanghai 200241, China
[10]Institute of Optoelectronics, Fudan University, Shanghai 200438, China

†These authors contributed equally to this work.
*Correspondence and requests for materials should be addressed to C. Z. (E-mail: zhangcheng@fudan.edu.cn) & X. Y. (E-mail: xyuan@lps.ecnu.edu.cn)

# Topological Lifshitz transition and one-dimensional Weyl mode in HfTe$_5$

**Abstract**

Landau band crossings typically stem from the intra-band evolution of electronic states in magnetic fields and enhance the interaction effect in their vicinity. Here in the extreme quantum limit of topological insulator HfTe$_5$, we report the observation of a topological Lifshitz transition from inter-band Landau level crossings using magneto-infrared spectroscopy. By tracking the Landau level transitions, we demonstrate that band inversion drives the zeroth Landau bands to cross with each other after 4.5 T and forms one-dimensional Weyl mode with fundamental gap persistently closed. The unusual reduction of the zeroth Landau level transition activity suggests a topological Lifshitz transition at 21 T which shifts the Weyl mode close to Fermi level. As a result, a broad and asymmetric absorption feature emerges due to the Pauli blocking effect in one dimension, along with a distinctive negative magneto-resistivity. Our results provide a strategy for realizing one-dimensional Weyl quasiparticles in bulk crystals.

**Main text**

In magnetic fields, electrons in crystals undergo cyclotron motion and transform the energy bands into discrete Landau levels. The formation of Landau levels gives rise to various phenomena such as Shubnikov–de Haas oscillations, integer and fractional quantum Hall effect, as well as composite fermions[1–3]. In two-dimensional (2D) systems like graphene and quantum wells, these Landau levels are non-dispersive. When additional energy such as spin or valley splitting exceeds the cyclotron energy, Landau levels meet with each other at critical magnetic fields and electron interaction is resultantly enhanced[4]. This type of intra-band Landau level crossings, which comes from the overlap of energy levels without dispersion, has been widely investigated and serves as an essential tool to modulate and analyze the Landau level spectrum[5–8]. In contrast, Landau levels in three-dimensional (3D) systems change into Landau bands due to the dispersion along the magnetic field direction. As the field varies, the evolution of the cyclotron energy and band splitting potentially lead to the Landau band crossings only at discrete momentums. Now, if considering an inverted-band system such as a weak topological insulator (TI), the spin splitting drives the lowest (zeroth) Landau bands of conduction and valence bands moving toward each other and finally crossing above a critical magnetic field at discrete momentums. Hereafter, the "inter-band Landau level crossing" denotes the crossing between conduction and valence Landau levels in momentum space. Notably, such inter-band Landau level crossings persistently close the fundamental band gap rather than the Landau gap alone. The further increased magnetic field only shifts the crossing momentums but keeps the band gap closed.

On the other hand, band crossing in momentum space generates quasiparticles which draw enormous research interests. The most mentioned examples involve Dirac equation which can be simplified to two massless Weyl equations[9] as $i\partial_t \psi_\pm = \pm p \psi_\pm$, where $\pm$ defines the chirality. By breaking either inversion or time reversal symmetry and forming band crossing, Weyl fermion has been theoretically proposed and experimentally realized[10–14]. It offers a platform to study the chiral fermion and leads to the discovery of chiral anomaly and other unique electromagnetic response[15–24]. The discussed inter-band Landau level crossings from TI lead to the effective one-dimensional (1D) structure without geometry confinement, as the magnetic field erases the in-plane dispersion. It mimics both the electronic structure and spin texture of the Weyl nodes formed by the Bloch band crossing which is defined as "1D Weyl mode" in the following. However, the effective Zeeman energy is generally much smaller than the band gap and the Fermi energy. Hence, the inter-band Landau level crossings, as well as the corresponding 1D Weyl mode in the quantum limit remain largely unexplored.

Here we report evidence of topological Lifshitz transition from inter-band Landau level crossings in the topological insulator $HfTe_5$. Owing to the low Fermi energy, $HfTe_5$ reaches the quantum limit in a very low field of ~ 1.5 T. A series of Landau level resonances along with band splitting behavior are revealed by magneto-infrared spectroscopy. By further ramping the magnetic field, we observe a highly unusual reduction of optical activity from zeroth Landau level transitions in the extreme quantum limit, which indicates the topological Lifshitz transition and the formation of 1D Weyl mode near the Fermi level. The electromagnetic response of this induced Weyl mode is revealed by both high-field optical and transport approaches, from which signatures of 1D Pauli blocking and negative magneto-resistivity are detected. The field-induced origin manifests the Weyl mode with immense density of states (DOS) near Fermi level, in stark contrast with the vanishing DOS in 3D Weyl node.

**Field-driven topological phase transitions and 1D Weyl mode**

Considering a 3D weak TI with an in-plane inverted gap, a series of magnetic-field-driven phase transitions are proposed as shown in Fig. 1. Applying magnetic field firstly leads to the formation of 1D Landau bands, which are parabolically dispersed along the field direction especially around the band edge at zero momentum. Once reaching the quantum limit at $B_0$, the Fermi level only crosses with the zeroth Landau band. By increasing magnetic fields, band inversion drives the two zeroth Landau bands moving toward and eventually forming inter-band crossings at $B_1$. At first, the Fermi level remains higher than the Lifshitz transition energy, so the system behaves as a trivial metal. Above $B_2$, a Lifshitz transition takes place accompanied by a topological phase transition with the Fermi surface divided into two parts with opposite spin textures. Such topological Lifshitz transitions originate from the dispersive Landau bands which are different from those in the Bloch bands[25]. The 1D crossing analog to the Weyl node formed by Bloch band crossing persists at higher magnetic fields. The overall phase transitions and main physics can be modeled by ideal low-energy Hamiltonian[26]

$$H(\mathbf{k}) = \hbar v_{Fx} k_x \tau_x \sigma_z - \hbar v_{Fy} k_y \tau_y \sigma_0 + \hbar v_{Fz} k_z \tau_x \sigma_x + \left[\Delta + M(k_x^2 + k_y^2) + M_z k_z^2\right] \tau_z \sigma_0, (1)$$

where $\mathbf{k}$ is the momentum, $\tau$ and $\sigma$ are Pauli matrix acting on the orbital and spin degrees of freedom, respectively. Band parameters include energy gap $2\Delta$, Fermi velocity $\mathbf{v_F}$ and band inversion parameters $M$, $M_z$. Fermi velocity and band inversion parameters also act as linear and parabolic contributions of energy dispersion[27]. Influence from additional perturbation terms[28–30] such as spin-orbit coupling and inversion symmetry breaking are further discussed in Supplementary Section III.

**Material realization and anisotropic band structure**

Among various TI candidates, we find $HfTe_5$ meets the material criteria to realize the above proposal. Similar to $ZrTe_5$, $HfTe_5$ locates at the boundary between weak TI and strong TI[31–33] with the electronic structure and band topology sensitive to the b-axis lattice constant. The access to the extreme quantum limit has led to various intriguing phenomena in $ZrTe_5$ and $HfTe_5$[34–39]. Photoemission and optical experiments reveal a temperature-dependent Fermi level and a TI phase at low temperature[40,41]. Figs. 2a-d exhibit the quantum oscillations from our transport measurement. A large positive magneto-resistivity is observed with a small Fermi vector $k_{ac} = 5.9 \times 10^{-3} \text{Å}^{-1}$, in agreement with the previous reports[42]. The fitted quantum limit in our sample is around $B_0 = 1.5 \pm 0.1$ T. Assuming an ellipsoid Fermi surface[42], we can obtain the cyclotron mass from temperature-dependent oscillation amplitude, as $m_a = 0.016\, m_e$, $m_b = 1.2\, m_e$, $m_c = 0.028\, m_e$ where $m_e$ is the free electron mass (refer to Supplementary Section I&II for details). Small $m_a$ and $m_b$ suggest quasi-linear in-plane dispersion, while $m_b$ indicates a parabolic dispersion along $k_b$ (Fig. 2d). Hence, we set the $v_z$ term to be zero with $M_z$ remains finite for HfTe$_5$[37]. With positive Δ, the in-plane band inversion of the TI requires $M < 0$. As shown later, the sign of $M_z$ determines the presence of inter-band Landau level crossings and Lifshitz transition. The subscripts x, y, z correspond to the a, c, b crystal axis of the HfTe$_5$. The a-c plane is treated isotropic because the cyclotron motion averages the in-plane response. The low carrier concentration along with the special anisotropic dispersion and band inversion serve as the prerequisites to observe the topological Lifshitz transition within a magnetic field of 35 T.

**High-field magneto-infrared spectrum**
The evolution of Landau bands is detected by magneto-infrared spectroscopy with optical transmittance $T_B/T_0$ (Fig. 2e) measured at the a-c plane of HfTe$_5$ under magnetic fields applied along b-axis (Faraday geometry). Here $T_B$ and $T_0$ are transmittance measured in magnetic field $B$ and zero field, respectively. A series of absorption peaks develop and evolve with magnetic fields on relative magneto-absorbance $A_B = -\ln(T_B/T_0)$ as shown in Fig. 2f. For those optical transitions labeled as $T_1$, $T_2$, $T_3$ …, the transition energy approximately follows $\omega \propto (\sqrt{n+1} + \sqrt{n})\sqrt{B}$ with $n = 1,2,3$ ..., which indicates a Dirac-type band in HfTe$_5$. Since the joint density of states diverges at the band edge, we focus on the $k_z = 0$ case at this stage and start from the optical transitions with the non-zeroth Landau band described by Eq. (2). The Landau bands are labeled as $L_n^s$ with $n = \pm 0, \pm 1, \pm 2$ ... and $s = \pm 1$ denoting the Landau index and spin index, respectively. Note the zeroth ($n = 0$) Landau bands are fully spin-polarized, distinct from all others. Band inversion leads to a field-dependent energy shift, which is most prominent for the zeroth Landau bands. Fig. 2g presents a schematic plot of Landau band edge energy $E(k_z = 0)$ versus $B$. The red and blue lines in Fig. 2g and Fig. 3c denote spin-polarized Landau bands. The optical transitions between Landau bands are restricted by a selection rule of $\Delta|n| = \pm 1$ for the ideal model. The photon energy for the $T_n$ transition ($L_{-n} \to L_{n+1}$ and $L_{-(n+1)} \to L_n$) follows Eq. (5) in the Method section, by which the transition index is assigned accordingly. The extracted Fermi velocity with $v_{ac} = 4.58 \times 10^5$ m/s is consistent with transport measurement. The observed inter-Landau-level transitions give the same intercept at zero field, corresponding to a small band gap of ~ 5 meV.

For optical transitions with energy lower than $T_1$ in Fig. 2f, it is readily seen to arise from the zeroth Landau bands. Here we refer to them together as $T_0$. As the Fermi level drops below the band edge of $L_{+1}$ at $B_0 \sim 1.5$ T, only the zeroth Landau bands are occupied so that the strength of $T_0$ transition is much higher than $T_1$ as shown in Panel i of Fig. 3a. When further increasing the magnetic field, $L_{+0}^-$ and $L_{-0}^+$ come closer owing to band inversion and eventually touch at the critical field $B_1$. It results in the splitting of $T_\alpha$, $T_\beta$ as shown by Panel ii and iv of Fig. 3a and the black arrow in Fig. 2f. The presence of small spin-orbit coupling mixes the spin of Landau band[20,28,35] which enables the spin-flipped $T_\alpha$, $T_\gamma$ transition at $B > B_1$ and also explains the absence of $T_0$ splitting at $B < B_1$ as summarized in Table 1. To extract the $B_1$ value, we perform multi-peak Lorentz fitting on the magneto-absorbance spectrum and extract the peak position of $T_\alpha$ and $T_\beta$ as shown in Panel iv of Fig. 3a. The splitting feature is confirmed and agrees with the model given by the solid lines. The intersecting magnetic field gives $B_1 \sim 4.5$ T. Details of $T_0$ transitions including the intensity distribution, field-dependent spin mixing, and the multi-peak Lorentz fitting are given in Supplementary Section III&XI. It is worthy to note that the energy of intra-band transition $L_{+0}^- \to L_{+1}^-$ is close to the inter-band transition $L_{-0}^+ \to L_{+1}^+$ after the system reaches the quantum limit, therefore they merge in $T_0$. Previous studies[35,43] have discussed a band edge touching picture of Landau levels in ZrTe₅ using a strong TI model or assuming a fixed Fermi energy. Their model gives a distinct $k_z$-dispersion and Fermi level variation as discussed below.

We note that the position of Fermi level is vital to determine the optical activity of the transitions beyond the quantum limit for HfTe₅. In 2D systems, the Landau levels are non-dispersive so that the Fermi level ultimately stays exactly at the lowest level of occupied bands. However, for 3D systems, the dispersion along the field direction of Landau bands may shift Fermi level away from the Landau bands energy at $k_z = 0$. In a 3D trivial insulator, the Fermi level always stays higher than the $L_{+0}$ but gradually converge to its band extrema due to the increasing DOS with magnetic fields (Fig. 3b, left panel). For a strong TI, the Fermi level first decreases with magnetic fields followed by an upturn (Fig. 3b, middle panel), since the system becomes fully gapped again after the band edge touching. As for a 3D weak TI, the Fermi level continues to drop after the band edge touching due to the persistent gap closure (Fig.3b, right panel). The dispersion of zeroth Landau bands along $k_z$ can be directly obtained from the proposed Hamiltonian $E_0^s(k_z) = -s(\Delta + M/l_B^2) + M_z k_z^2$. The effect of Fermi level shifting with fields is mostly overlooked in previous studies of ZrTe₅/HfTe₅ Landau level spectrum. Below we will show that Landau level transitions in HfTe₅ can be well explained by this picture.

Generally speaking, the optical transition activity of $T_0$ persists beyond the quantum limit and the intensity grows with fields, since the Fermi level no longer crosses any Landau band edge. However, as indicated by the bottom orange arrow in Fig. 2f, $T_\alpha$ becomes weakened and gradually disappears above $B_2 \sim 21$ T while all other $T_0$ transitions get enhanced. Panels ii, iii, and v in Fig. 3a also clearly indicate the reduction of $T_\alpha$. This striking phenomenon contrasts the traditional argument and suggests that the Fermi level further crosses $L_{-0}^+$ around $B_2$. Above $B_2$, the emptying of $L_{-0}^+$ forbids all transitions initializing from $L_{-0}^+$ (Fig. 3c), hence $T_{\alpha 1}$ vanishes accordingly. In contrast, $T_\beta$ and $T_\gamma$ persist above $B_2$. Fig. 3c shows the assigned transitions for $T_\beta$ and $T_\gamma$. While $T_{\beta 1}$ (solid dark green) maintains the optical activity through $B_2$, $T_{\beta 2}$ (dashed dark green) become active, resulting in the increased overall intensity of $T_\beta$ above $B_2$. In comparison, $T_{\beta 1}^*$ (solid light green) experiences similar variation as $T_{\alpha 1}$, but locates at the same energy as $T_\beta$. Considering the spin conservation nature and the overlapping of $T_{\beta 1,2}$ and $T_{\beta 1}^*$, $T_\beta$ shows the highest intensity among $T_0$. The remaining $T_\gamma$ naturally comes with the highest energy of the transition $L_{+0}^- \to L_{+1}^+$. With a similar argument, while $T_{\gamma 1}$ remains active through $B_2$, $T_{\gamma 2}$ transition is expected to appear only after $B_2$, where the increased intensity and broadening of $T_\gamma$ are observed (the upper orange arrow in Fig. 2f). Due to the level broadening effect from disorder and finite temperature, the discussed optical activity variation will not change abruptly but gradually fade away. To quantitatively analyze the optical activity variation of $T_\alpha$, we perform the multi-peak Lorentz fitting to extract the Pauli-blocking-induced normalized spectral weight ($SW_{PB}$) variation as shown in Panel v of Fig. 3a (fitting details and model predictions are presented in Supplementary Section III&XI). The $SW_{PB}$ of $T_\alpha$ is expected to drop most steeply around $B_2$. Therefore, the extreme point of its first derivative gives the critical field $B_2 \sim 21$ T. Meanwhile, fitted peak width is found to increase with magnetic fields, indicating the presence of impurity scattering [44–46].

These observations suggest that the as-grown HfTe$_5$ is indeed a 3D Weak TI. The experimental data can be well fitted by the $\mathbf{k} \cdot \mathbf{p}$ model with the band parameters given in Supplementary Section IV. To further verify the consistency between the experimental results and our model, we reproduce the magneto-optical-conductivity spectrum with fitting parameters as shown in Fig. 3d. The observed spectrum is well reproduced as well as the overall topological phase transitions throughout magnetic field range. The optical activity of $T_0$ transition is summarized in Table 1. We also discuss the discrepancy between ZrTe$_5$ and HfTe$_5$ in Supplementary Section V.

After the topological Lifshitz transition at $B_2$, two spin-polarized Landau bands cross near the Fermi level, analog to 3D Weyl node formed by Bloch band crossing. Under high magnetic fields, the low-energy excitation resembles effective 1D Weyl quasiparticles because the Landau bands only disperse along $k_z$. Moreover, the Fermi velocity of the obtained 1D Weyl mode can be tuned from $5 \times 10^4$ m/s at $B_2 \sim 21$ T to $7 \times 10^4$ m/s at the highest experimental field of 35 T. The theoretically allowed Fermi velocity is expected to reach $\sim 2 \times 10^5$ m/s before annihilating the Weyl points at the boundary of Brillouin zone around $\sim 300$ T (estimation details are given in Supplementary Section VI&VIII).

**Optical and transport response of 1D Weyl mode**
The field-induced 1D Weyl mode also features high DOS near the Fermi level. We first compare it with conventional 3D Weyl node in Fig. 4a. By tuning the Fermi level through doping, the DOS near Fermi level vanishes at the 3D Weyl node position (Fig. 4a, Panel i). In HfTe$_5$, the DOS of each Landau band increase with magnetic fields. As a result, higher magnetic fields push the Fermi energy toward the crossing point and increase the nearby DOS which eventually diverges around the charge neutral point (CNP) as shown in Panel ii of Fig. 4a.

1D feature of the Weyl mode also gives rise to distinct electromagnetic responses compared to the one in other dimensions. For systems with linear dispersion, it is well-known that the real part of optical conductivity follows $\sigma_1 \propto \omega^{d-2}$, where $d$ is dimension (Fig. 4a, Panel iii and iv). Pauli blocking effect resembles the onset of this behavior at twice the Fermi energy. It has been fully verified in 2D and 3D systems such as graphene and $Cd_3As_2$[47,48]. For 3D Weyl node at finite temperature and scattering conditions, the Pauli blocking effect only mildly changes the conductivity spectrum (blue dashed line). In 1D case, however, the divergence of optical conductivity yields a peak feature, which is exactly the case of $HfTe_5$ under magnetic fields (blue and green dashed line). From 1D Pauli blocking under finite temperature and scattering conditions (Fig. 4b), one can predict (1) a peak on optical conductivity with high-energy tail, (2) peak height increases with fields owing to the higher DOS of Landau bands, (3) comparatively stable frequency with magnetic fields, (4) appearance of peak feature after $B_2$. Apart from the discussed spin mixing effect in $HfTe_5$, the transition between two zeroth Landau levels is allowed by the orbit mixing effect (refer to Supplementary Section III for quantitative verification) from the inversion symmetry breaking at low temperatures[30]. The complete selection rules including $\Delta|n| = 0, \Delta s = \pm 2$ and the quantitative derivation are given in Supplementary Table S1 and Section III. In experiment, we observe a newly formed set of peaks $T_w$ with a prominent high-energy tail above $B_2$ as shown in Fig. 2f and denoted by the black arrow in Fig. 4d, whose peak height increases with fields. The presence of $T_w$ near $B_2$ suggests the relation with the induced 1D Weyl mode but the low frequency distinguishes it from $T_0$ transition. The typical features of $T_w$ fit well with the 1D Pauli blocking behaviors discussed above.

For 1D Weyl fermion, chiral anomaly takes a simple form of $d(n_{R/L}^{1D})/dt = \pm eE/h$ where $n_{R/L}^{1D}$ denotes the number of right-handed and left-handed 1D Weyl fermions (Fig. 4c). Here $E$ is the external electric field. The chiral anomaly essentially describes a 1D conductivity channel. To verify the influence of the topological Lifshitz transition on the electrical property of $HfTe_5$, we measure the electrical resistivity along the a-and b-axis at pulsed magnetic fields, respectively. The magnetic field is applied along the b-axis. As shown in Fig. 4e, $R_{xx}$ exhibits prominent linear behavior before saturating at high fields, which may result from the impurity scattering characterized by the screened Coulomb potential (refer to Supplementary Section X for more discussion)[49,50]. In contrast, $R_{zz}$ experiences a clear drop around 20 T, which is reminiscent of chiral-anomaly-induced negative magneto-resistivity in 3D Weyl semimetals. The critical field for negative $R_{zz}$ is close to the field of topological Lifshitz transition ($B_2 \sim 21$ T). Above this field, the 1D chiral anomaly gives rise to the b-axis conductive channel with chiral current proportional to $E$. Recall that the magnetic field increases the Landau bands DOS near the Fermi level (Fig. 4a panel ii), resulting in chiral current proportional to magnetic fields. Consequently, field-induced 1D quasiparticles in HfTe$_5$ present chiral anomaly with an $E \cdot B$-like characteristic similar to the 3D counterpart, which explains the anomalous decrease of $R_{zz}$ above $B_2$ compared with $R_{xx}$.

Different from symmetry-protected topological phases such as 3D Dirac and Weyl semimetals, the 1D Weyl mode in HfTe$_5$ is not symmetry-protected. Based on our magneto-infrared spectroscopy results, the gap size should be extremely small as a few meV with negligible influence in our experimental regime. It is also worthy to note that the formation of 1D Weyl mode is essentially much more difficult in strong TI ($M < 0$, $M_z < 0$). Without spin-orbit coupling, applying quantum limit magnetic fields in strong TI leads to the cross of zeroth Landau bands before $B_1$ and a gap opening after that. Therefore, dropping the Fermi level through $L_{-0}$ before $B_1$ and realizing 1D Weyl mode near Fermi level are not accessible as shown in Fig. 3b. HfTe$_5$, in our case, features in-plane band inversion ($M < 0$) but an out-of-plane trivial gap ($M_z > 0$) as a weak TI. The phase transition at $B_1$ changes the in-plane gap to be trivial but, most importantly, turns the out-of-plane direction into band inversion state, which guarantees a Weyl mode after $B_2$. The observed Landau band Lifshitz transition can be further detected by other spectroscopic techniques such as scanning tunneling microscopy under high magnetic field which could further verify the 1D Weyl mode and explore the electromagnetic response in real space.

**Conclusion**

In summary, we report the magneto-infrared spectroscopic evidence of field-induced topological Lifshitz transition and 1D Weyl mode in HfTe$_5$. In magnetic fields, band inversion results in the inter-band Landau level crossings accompanied by the persistent closure of band gap. With Fermi level further dropping with fields, the reduction in optical transitions indicates the presence of topological Lifshitz transition and 1D Weyl mode near the Fermi level. The observed 1D Pauli blocking behavior and negative magneto-resistivity agree well with the electromagnetic response of 1D Weyl fermions. The overall magneto-infrared features and field-driven phase transitions can be quantitatively explained by the TI $\mathbf{k} \cdot \mathbf{p}$ model. The realization of 1D Weyl mode from inter-band Landau levels establishes a unified strategy of topological phase engineering.

**Acknowledgements**

X.Y. was supported by the National Natural Science Foundation of China (Grant No. 12174104, No. 62005079 and No. 62111530237), the Shanghai Sailing Program (Grant No. 20YF1411700), the International Scientific and Technological Cooperation Project of Shanghai (Grant No. 20520710900), and a start-up Grant from East China Normal University. C.Z. was supported by the National Natural Science Foundation of China (Grant No. 12174069), Shanghai Sailing Program (Grant No. 20YF1402300), Natural Science Foundation of Shanghai (Grant No. 20ZR1407500), the young scientist project of MOE innovation platform and the start-up Grant at Fudan University. H.-Z.L. was supported by the National Natural Science Foundation of China (11925402). A portion of this work was performed at the National High Magnetic Field Laboratory, which is supported by National Science Foundation Cooperative Agreement No. DMR-1644779 and the State of Florida. Part of the sample fabrication was performed at Fudan Nano-fabrication Laboratory. We thank Milan Orlita, Marek Potemski, Hugen Yan, Yuxuan Jiang, Zhi-Guo Chen, Zhenrong Sun, Chun-Gang Duan, Fangyu Yue, Bobo Tian and Yanwen Liu for the helpful discussion.

**Author contributions**
X.Y. conceived the idea and supervised the overall research. X.M., B.L. and Y.M. carried out the growth of the $HfTe_5$ single crystals. M.O., W.W., Z.S., Y.D. performed the magneto-infrared experiments. C.Z. and Y.W. conducted magneto-transport experiments. W.W., X.Y., Z.Y., F.Q., and H.-Z.L. performed the theoretical analyses based on the $\mathbf{k}\cdot\mathbf{p}$ model. X.Y., W.W., C.Z., Z.S., and J.C. wrote the paper with the help of all other co-authors.

**Competing interests**
The authors declare no competing interest.

# Table

**Table 1 | Optical transition activity at different field-driven phases**

| Transition | $T_{\alpha 1}$ $L_{-0}^{+} \to L_{+1}^{-}$ | $T_{\alpha 2}$ $L_{-1}^{+} \to L_{+0}^{-}$ | $T_{\beta 1}$ $L_{+0}^{-} \to L_{+1}^{-}$ | $T_{\beta 2}$ $L_{-1}^{+} \to L_{-0}^{+}$ | $T_{\beta 1}^{*}$ $L_{-0}^{+} \to L_{+1}^{+}$ | $T_{\beta 2}^{*}$ $L_{-1}^{-} \to L_{+0}^{-}$ | $T_{\gamma 1}$ $L_{+0}^{-} \to L_{+1}^{+}$ | $T_{\gamma 2}$ $L_{-1}^{-} \to L_{-0}^{+}$ |
|---|---|---|---|---|---|---|---|---|
| $0 \sim B_0$ | × | × | × | × | × | × | × | × |
| $B_0 \sim B_1$ | × | × | √ | × | √ | × | × | × |
| $B_1 \sim B_2$ | √ | × | √ | × | √ | × | √ | × |
| $B_2$ | × | × | √ | √ | × | × | √ | √ |

# Figures

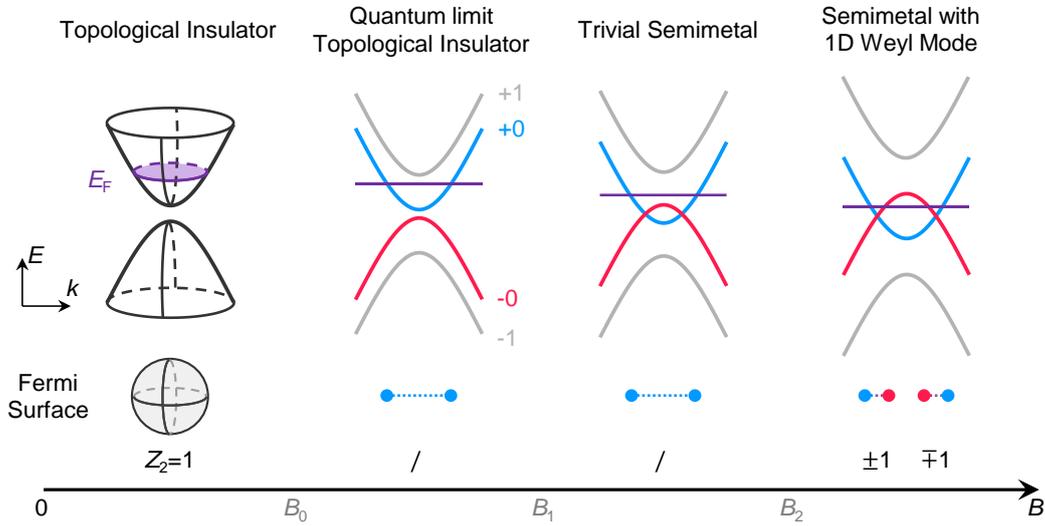

**Fig. 1 | Schematic plot of proposed magnetic-field-driven phase transitions.** Considering a 3D weak TI at zero field with $\Delta > 0$, $M < 0$, $M_z > 0$. The Landau quantization of the TI features both band inversion and full spin polarization of the zeroth Landau bands denoted by red and blue lines, while the black and gray lines denote original energy bands and high-index Landau bands. After reaching the quantum limit at $B_0$, the Fermi level only crosses the zeroth Landau band. Characteristic band inversion leads to the crossing of zeroth Landau bands after critical field of $B_1$. With Fermi level staying high, the system still behaves as a trivial semimetal until the Lifshitz transition at $B_2$. Fermi surfaces experience a splitting, which is accompanied by a topological transition where zeroth Landau bands form effective 1D and spin-polarized band crossing near Fermi energy.

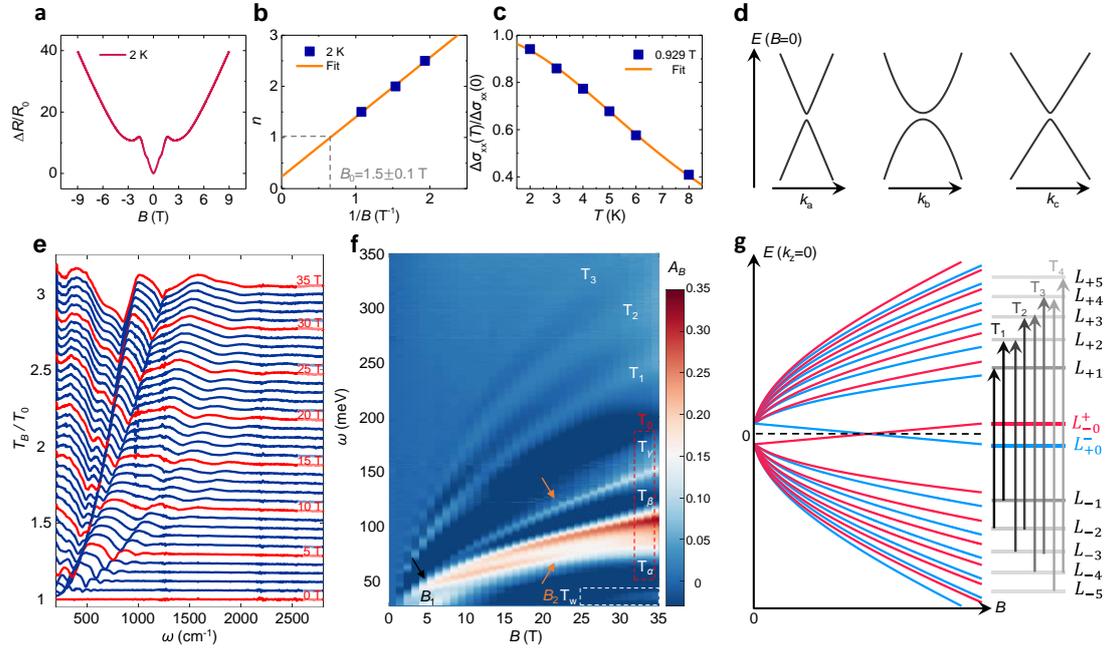

**Fig. 2 | Band structure and magneto-infrared spectroscopy in HfTe$_5$. a**, Magneto-resistivity with large ratio and quantum oscillation. **b**, Fan diagrams exhibiting small Fermi surface and quantum limit at $B_0 \sim 1.5$ T. **c**, Temperature-dependent oscillation amplitude. **d**, The conclusive anisotropic band structure of HfTe$_5$ for model parameter setting. **e**, Relative magneto-transmittance spectra $T_B/T_0$ at different magnetic fields. All curves are vertically stacked for clarification. The Landau level transitions (dips) systematically evolve with magnetic fields. **f**, False-color plot of the magneto-absorbance $A_B = -\ln(T_B/T_0)$. The assignments of Landau level transitions are labeled in T$_n$. The black arrow points to the splitting features originating from the zeroth Landau band edge touching at critical field of $B_1$. The orange arrows present the optical activity variation due to the Lifshitz transition and resultant formation of Weyl mode near Fermi level at critical field of $B_2$. The white dashed box exhibits the optical features from the 1D Pauli blocking effect. **g**, The schematic Landau band edge energy of HfTe$_5$ under magnetic fields. The arrows exhibit the non-zeroth Landau level transitions.

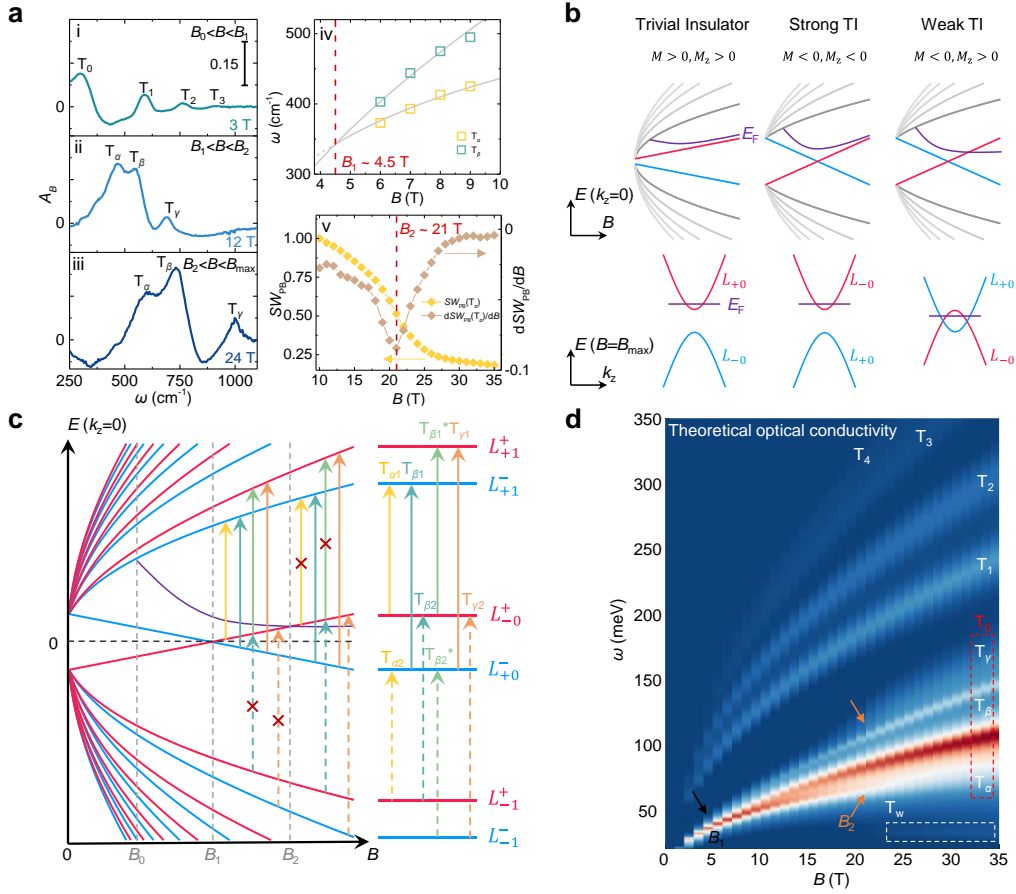

**Fig. 3 | Optical activity of Landau level transitions and field-driven phase transitions. a**, Relative magneto-absorbance spectrum at magnetic field regime (i) after reaching the deep quantum limit at $B_0 < B < B_1$, showing $T_0$ as the most prominent transition with only the zeroth Landau band occupied; (ii) after the touching of zeroth Landau band edge at $B_1 < B < B_2$, showing the split $T_\alpha$ and $T_\beta$ indicating the crossing of zeroth Landau bands; (iii) after Lifshitz transition at $B > B_2$, showing the reduction of $T_\alpha$ in contrast with the continuous intensity boost of $T_\beta$ and $T_\gamma$ suggesting the emergence of 1D Weyl mode near Fermi level. The fitted peak position in Panel (iv) confirms the splitting behavior, which agrees with the model (solid line), and gives $B_1 \sim 4.5$ T. The normalized Pauli-blocking-induced spectral weight $SW_{PB}$ of $T_\alpha$ and its first derivative in Panel (v) prove the reduction of $T_\alpha$ and give $B_2 \sim 21$ T. **b**, Landau bands and Fermi level variation with magnetic fields in small gap insulators with different topology. In general cases, Femi level does not cross any Landau band extrema after the quantum limit. An exception is presented in weak TI where Fermi level (in purple) crosses $L_{-0}$ which explains the unusual reduction of $T_\alpha$ and 1D Weyl mode. **c**, Landau band extrema energy versus magnetic field focusing on $T_0$ transition. The activity of optical transition is plotted by arrows with and without red crossing. $T_\alpha$ in yellow experiences activity reduction through $B_2$ while $T_{\gamma2}$ experience the opposite. **d**, Model prediction of the real part of magneto-optical-conductivity spectrum based on the fitting parameters which reproduce the experimental features of inter-Landau level transitions.

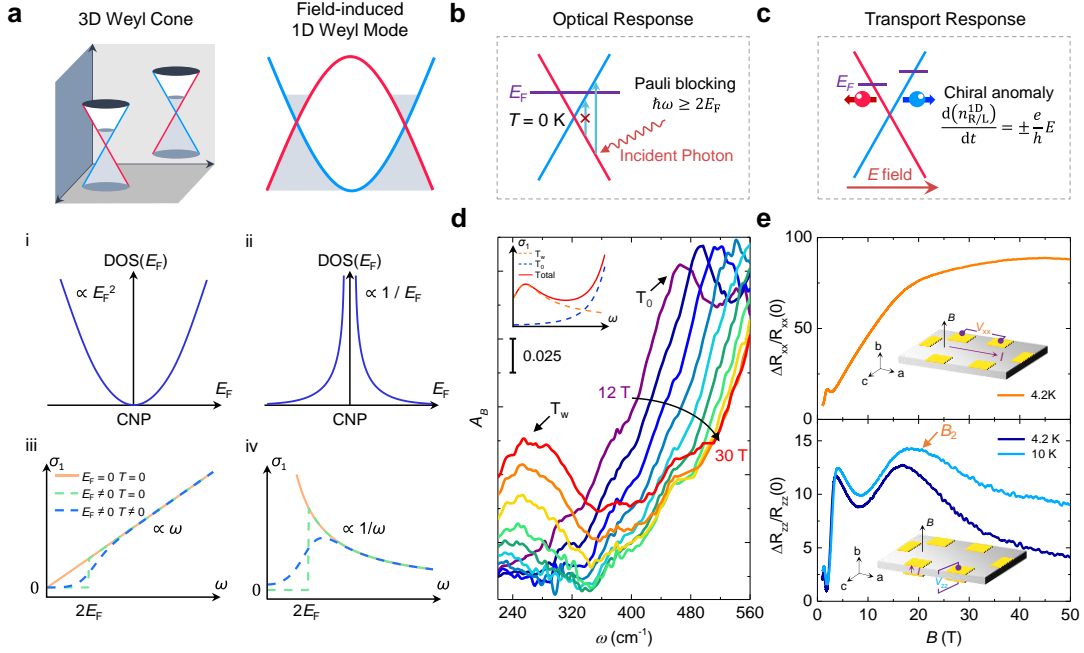

**Fig. 4 | Signature of 1D Pauli blocking and chiral anomaly. a**, Comparison between 3D Weyl cone and field-induced 1D Weyl mode. The upper, middle and lower panels denote the comparison of band structure, the DOS near Fermi level, and optical conductivity with Pauli blocking, respectively. The field-induced nature of 1D Weyl mode results in high DOS near Fermi level contrasting the vanishing DOS in 3D Weyl node. $\sigma_1 \propto \omega^{d-2}$ ($d$ denotes the dimension) are expected for linear band, therefore, predicting a peak feature in 1D system. **b**, **c**, Schematic plots of 1D Pauli blocking and 1D chiral anomaly as optical and electrical response of 1D Weyl mode, respectively. **d**, The optical conductivity at different magnetic fields. The appearance of $T_w$ after $B_2$ with high energy tail, increasing oscillators strength, and frequency stability, consistent with 1D Pauli blocking. The inset shows a schematic spectrum for the Pauli blocking peak near the Landau level transition. **e**, Magneto-resistivity measurement. The upper panel presents the traditional in-plane Hall bar geometry. The lower panel presents the measurement with both $E$ and $B$ parallel to 1D Weyl mode direction, therefore, including the out-of-plane conductivity contribution after $B_2$.

## Methods
### Material choice and crystal preparation
HfTe$_5$ is chosen for realizing the proposed 1D Weyl mode due to low Fermi energy, accessible quantum limit and locating at the strong/weak TI phase boundary[51–62]. The HfTe$_5$ single crystals were prepared by the standard chemical vapor transport method. Stoichiometric mixtures of Hf and Te powder were sealed in an evacuated quartz tube. Iodine was added as the transport agent. The tube was placed in the two-zone furnace with a hot end temperature setting at 770 K and a temperature gradient of 100 K. After reaching the designed temperature, the condition is kept for two weeks. Needle-like single crystals are obtained after cooling down to room temperatures.

### Magneto-infrared measurement
Magneto-infrared spectroscopy was performed using an FT-IR spectrometer (Bruker IFS-66) with a 35 T resistive magnet at National High Magnetic Field Laboratory, Tallahassee. The collimated infrared beam from the spectrometer was propagating inside an evacuated beamline and focused at the top of the probe with a diamond window. Then the IR beam was guided through a brass light pipe to the sample space, which was cooled down to liquid helium temperature by a small amount of helium exchange gas. The thin HfTe$_5$ flakes with tens of microns thickness were mechanically exfoliated from bulk HfTe$_5$ single crystals. After mounting them on the transmission sample holder, DC magnetic field was applied along the crystallographic b-axis in Faraday geometry. The infrared beam went through the sample and then was detected by a 4.2 K composite silicon bolometer located just a short distance below. The single transmission spectra were collected with an acquisition time of about 3 min.

### Magneto-transport measurement
Low-field magneto-transport measurements were carried out using a superconducting magnet with standard Lock-in technique. High-field magneto-transport measurements were performed in a pulsed magnet up to 50 T.

### k · p model of HfTe$_5$
The ideal Hamiltonian of HfTe$_5$ is given by Eq. (1), when applying magnetic field $B$ along the z-axis, the energy of Landau bands in the TI reads

$$E_n^s(k_z = 0) = -s\frac{M}{l_B^2} + \alpha\sqrt{\left(\Delta + 2\frac{M}{l_B^2}|n|\right)^2 + 2\frac{\hbar^2 \bar{v}_F^2}{l_B^2}|n|}, \quad n = \pm 1, \pm 2, \pm 3 \ldots, \quad (2)$$

$$E_{n=0}^s(k_z = 0) = -s\left(\Delta + \frac{M}{l_B^2}\right), \quad (3)$$

$$E_{n=0}^s(k_z) = -s\left(\Delta + \frac{M}{l_B^2}\right) + M_z k_z^2, \quad (4)$$

where $\bar{v}_F = \sqrt{v_{Fx} v_{Fy}}$, $n$, $\alpha = \pm 1$ and $s = \pm 1$ denote Landau index, carrier type index, and spin index, respectively. $l_B = \sqrt{\hbar/eB}$ is the magnetic length. The optical transitions between Landau levels in the ideal case generally follow selection rules of $\Delta |n| = \pm 1$. Therefore, the optical transitions occur from both $L_{-n} \to L_{n+1}$ and $L_{-(n+1)} \to L_n$, where $L_n$ denotes the $n^{\text{th}}$ Landau level. The optical transitions at $k_z = 0$ require photon energy following

$$\omega(n,B) = \sqrt{\left[\Delta + 2\frac{M}{l_B^2}(|n|+1)\right]^2 + 2\frac{\hbar^2 \bar{v}_F^2}{l_B^2}(|n|+1)}$$
$$+ \sqrt{\left(\Delta + 2\frac{M}{l_B^2}|n|\right)^2 + 2\frac{\hbar^2 \bar{v}_F^2}{l_B^2}|n|}. \quad (5)$$

Detailed discussions including Zeeman effect[63], disorder[64–67], and other perturbation terms are provided in Supplementary Section III&X.

**Data availability**

Source data are provided with this paper. All other supporting data are available from the corresponding author upon reasonable request.